# Electronic structure and chemical bonding in Ti$_2$AlC investigated by soft x-ray emission spectroscopy

M. Magnuson[1], O. Wilhelmsson[2], J. -P. Palmquist[2], U. Jansson[2], M. Mattesini[3], S. Li[1], R. Ahuja[1] and O. Eriksson[1]

[1] *Department of Physics, Uppsala University, P. O. Box 530, S-751 21 Uppsala, Sweden*
[2] *Department of Materials Chemistry, The Ångström Laboratory, Uppsala University, P.O. Box 538 SE-75121 Uppsala*
[3] *Departamento de F'isica de la Tierra, Astronom'ia y Astrof'isica I, Universidad Complutense de Madrid, E-28040, Spain*

**Abstract**
The electronic structure of the nanolaminated transition metal carbide Ti$_2$AlC has been investigated by bulk-sensitive soft x-ray emission spectroscopy. The measured Ti *L*, C *K* and Al *L* emission spectra are compared with calculated spectra using *ab initio* density-functional theory including dipole matrix elements. The detailed investigation of the electronic structure and chemical bonding provides increased understanding of the physical properties of this type of nanolaminates. Three different types of bond regions are identified; the relatively weak Ti *3d* - Al *3p* hybridization 1 eV below the Fermi level, and the Ti *3d* - C *2p* and Ti *3d* - C *2s* hybridizations which are stronger and deeper in energy are observed around 2.5 eV and 10 eV below the Fermi level, respectively. A strongly modified spectral shape of the *3s* final states in comparison to pure Al is detected for the buried Al monolayers indirectly reflecting the Ti *3d* - Al *3p* hybridization. The differences between the electronic and crystal structures of Ti$_2$AlC, Ti$_3$AlC$_2$ and TiC are discussed in relation to the number of Al layers per Ti layer in the two former systems and the corresponding change of the unusual materials properties.

**Introduction**
Nanolaminated ternary carbides and nitrides, also referred to as MAX phases, denoted M$_{n+1}$AX$_n$, where n=1, 2 and 3 represents 211, 312 and 413 crystal structures, respectively, have recently been the subject of intense research [1-3]. M denotes an early transition metal, A is a p-element, usually belonging to the groups IIIA and IVA, and X is either carbon or nitrogen [4]. These layered materials exhibit an unusual and unique combination of metallic and ceramic properties, including high strength and stiffness at high temperatures, resistance to oxidation and thermal shock, as well as high electrical and thermal conductivity [5]. The macroscopic properties are closely related to the underlying electronic structure, the crystal structure of the constituent elements and their monolayers. Generally, the MAX-phase family has a hexagonal crystal structure with near close-packed layers of the M-elements interleaved with square-planar slabs of pure A-elements, where the X-atoms fill the octahedral sites between the M-atoms. The A-elements are located at the center of trigonal prisms that are larger than the octahedral X sites. The difference between the 211, 312 and 413 structures is the number of `inserted' A-monolayers per M layer. The A/M ratios are 0.5, 0.33 and 0.25 for the 211, 312 and 413 structures, respectively. The 312 and 413 structures are more complicated than the 211 structure with two different M sites, denoted $M_I$ and $M_{II}$. The 413 structure also has two different X sites, denoted $X_I$ and $X_{II}$.

The history of the 211-crystal structure dates back to the early 1930's when these materials were referred to as Hägg phases with a large group of energetically stable variants [6]. Although the history of *MAX* phases is quite long, the recent improvements in synthetization processes has led to a renaissance of these compounds due to the discovery of the unique mechanical and physical properties [5,7]. The Ti-Al-C system is the most important and stable set of MAX phases due to excellent oxidation resistance at high temperature above 1100º C. Insertion of Al monolayers into a TiC matrix implies that the strong Ti-C bonds are broken up and replaced by weaker Ti-Al bonds with a cost of energy. Thus, in $Ti_2AlC$, every second single monolayer of C atoms have been replaced by Al layers. The TiC layers surrounding the Al monolayers are then twinned with the Al layer as a mirror plane. Figure 1 shows the crystal structure of $Ti_2AlC$ (211) with the thermodynamically stable nanolaminates of binary Ti-C-Ti slabs separated by softer Ti-Al-Ti slabs with weaker bonds [8]. For comparison, the 312 crystal structure is also shown where there are two different Ti atoms, $Ti_I$ and $Ti_{II}$. As observed in Fig. 1, the 211 crystal structure contains $Ti_{II}$ atoms with chemical bonds both to the C and the A-atoms while the 312 structure also contains $Ti_I$ atoms which only bond to C. $Ti_2AlC$ is not only the most stable Ti-Al-C compound; it has a lower density than other MAX-phases with mechanical properties similar to $Ti_3AlC_2$ but is easier to machine in its bulk form. The elastic properties, such as Young's modulus (E), change with phase and composition i.e., $Ti_2AlC$ (240 GPa) is softer than $Ti_3AlC_2$ (260 GPa) which is even softer than the prototype compound $Ti_3SiC_2$ (320 GPa) [9]. The change of elastic properties with crystal structure is mainly related to the fact that the 211 structure contain a smaller part of the strong Ti-C bonds and thus generally exhibit more metallic-like attributes and softness compared to the 312 and 413 structures, which exhibit more carbide-like attributes. The weak Ti-Al bonds also affect the tribological properties, such as wear performance and friction [5]. The physical properties of crystallographically oriented thin films of MAX phases thus provide opportunities for particular industrial applications such as protective coatings, sliding/gliding electrical contacts and heating elements.

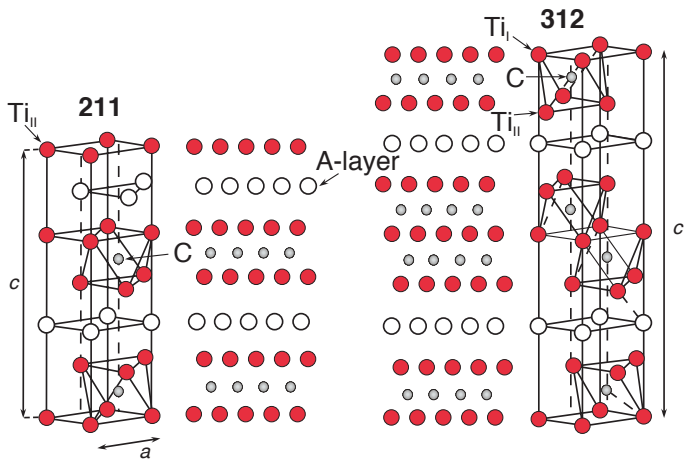

**Fig. 1:** The hexagonal crystal structures of 211 ($Ti_2AlC$) in comparison to 312. There is one Al layer for every second layer of Ti in $Ti_2AlC$. The $Ti_{II}$ atoms have chemical bonds to both C and Al while the $Ti_I$ atoms only bond to C. The lengths of the measured (calculated) *a* and *c*-axis of the unit cell of $Ti_2AlC$ are 3.04, (3.08) Å and 13.59, (13.77) Å, respectively.

Previous experimental investigations of the electronic structure of $Ti_2AlC$ include valence-band x-ray photoemission (XPS) [10,11]. However, XPS is a surface sensitive method, which is not element specific in the valence band. Theoretically, it has been shown by *ab initio* bandstructure calculations that there should be significant differences of the partial density-of-states (pDOS) of Ti, C and Al between different crystal structures [12-15]. In recent studies, we investigated the three 312 phases $Ti_3AlC_2$, $Ti_3SiC_2$ and $Ti_3GeC_2$ [16] and the 413 phase $Ti_4SiC_3$ [17]. In contrast to $Ti_3SiC_2$ and $Ti_3GeC_2$, a pronounced shoulder about 1 eV below the Fermi level was identified in the Ti $L_{2,3}$ soft x-ray emission (SXE) spectra of $Ti_3AlC_2$. From that study, it is clear that the physical and mechanical macroscopic properties of MAX phases can be further understood from detailed investigations of the underlying electronic structure, and in particular, the M-A and M-X chemical-bond interactions.

In the present paper, we investigate the electronic structure of $Ti_2AlC$, using bulk-sensitive and element-specific SXE spectroscopy with selective excitation energies around the Ti *2p*, C *1s* and Al *2p* thresholds. The SXE technique is more bulk sensitive than other electron-based spectroscopic techniques. Due to the involvement of both valence and core levels, the corresponding difference in energies of emission lines and their selection rules, each kind of atomic element can be probed separately. This makes it possible to extract both elemental and chemical information of the electronic structure. The SXE spectra are interpreted in terms of partial valence band density of states (pDOS) weighted by the transition matrix elements. The main objective of the present investigation is to study the nanolaminated internal electronic structure and the influence of hybridization among the constituent atomic planes in $Ti_2AlC$, in comparison to $Ti_3AlC_2$ and TiC with the aim to obtain an increased understanding of the physical and mechanical properties.

**Experimental**

**X-ray emission and absorption measurements**
The SXE and x-ray absorption spectroscopy (XAS) measurements were performed at the undulator beamline I511-3 at MAX II (MAX-lab National Laboratory, Lund University, Sweden), comprising a 49-pole undulator and a modified SX-700 plane grating monochromator [18]. The SXE spectra were measured with a high-resolution Rowland-mount grazing-incidence grating spectrometer [19] with a two-dimensional detector. The Ti *L* and C *K* SXE spectra were recorded using a spherical grating with 1200 lines/mm of 5 m radius in the first order of diffraction. The XAS spectra at the Ti *2p* and C *1s* edges were measured with 0.1 eV resolution. During the Ti *L* and C *K* SXE measurements, the resolutions of the beamline monochromator were 0.7, and 0.2 eV, respectively. The SXE spectra were recorded with spectrometer resolutions of 0.7 and 0.2 eV, respectively. All the measurements were performed with a base pressure lower than $5 \times 10^{-9}$ Torr. In order to minimize self-absorption effects [20], the angle of incidence was about 20° from the surface plane during the emission measurements. The x-ray photons were detected parallel to the polarization vector of the incoming beam in order to minimize elastic scattering.

**Deposition of the Ti$_2$AlC film**

Figure 2 shows θ−2θ diffractograms of the deposited TiC and Ti$_2$AlC films. The TiC$_x$(111) ($x \sim 0.7$, 2000 Å thick) and Ti$_2$AlC (000$l$) (5000 Å thick) films were epitaxially grown on Al$_2$O$_3$ (000$l$) substrates at 300° C and 900° C, respectively, by dc magnetron sputtering [21]. Elemental targets of Ti, C and Al, and a 3.0 mTorr Ar discharge were used. To promote a high quality growth of the MAX phase, a 200 Å thick seed layer of TiC$_{0.7}$(111) was initially deposited. For further details on the synthesis process, the reader is referred to Refs [22-24].

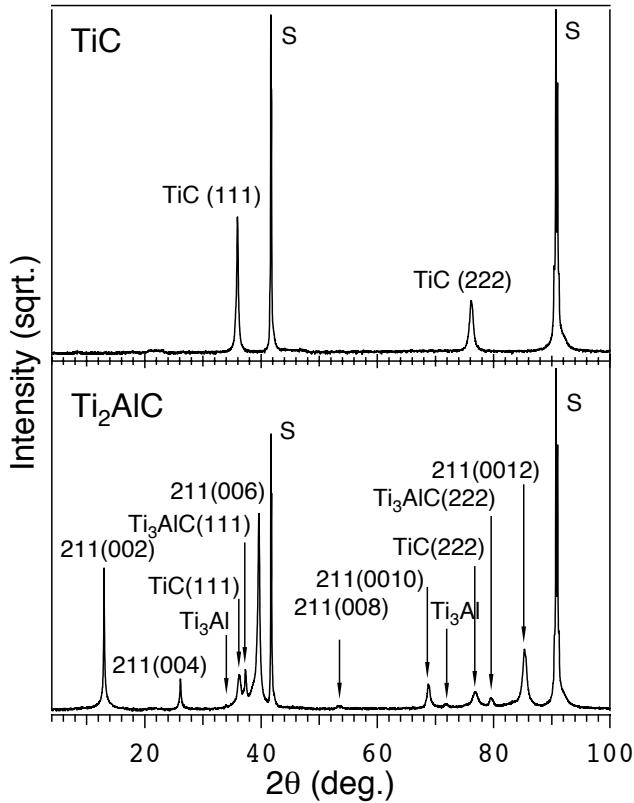

**Fig. 2:** Top, x-ray diffractogram of TiC. Bottom, x-ray diffraction from the Ti$_2$AlC sample. S denotes the contribution from the Al$_2$O$_3$ substrate. The TiC peaks in Ti$_2$AlC originates from the seed layer interface.

The two most intense peaks in the Ti$_2$AlC sample in Fig. 2 corresponding to α-Al$_2$O$_3$(0006) and α-Al$_2$O$_3$(0012) reflections, originate from the substrate. As observed, the other peaks mainly originate from Ti$_2$AlC(000$l$). Small contributions from Ti$_3$Al, Ti$_3$AlC (*lll*) and the TiC (*lll*) seed layer are also observed. The weak intensities of the Ti$_3$Al and Ti$_3$AlC peaks indicates that these phases only represent a minority phase and do not affect the x-ray emission measurements. The TiC seed layer does not either influence the x-ray emission measurements since the probe depth is less than 2000 Å at 20 degrees incidence angle. The relatively low intensities of the additional peaks show that the film mainly consists of single-phase *MAX*-material. Furthermore, the fact that the diffractogram shows only Ti$_2$AlC of {000$l$}-type suggests highly textured or epitaxial films. X-ray pole figures verified that the growth indeed was epitaxial, and determined the relation to Ti$_2$AlC(000$l$)//TiC(111)//Al$_2$O$_3$(000$l$) with an in-plane orientation of Ti$_2$AlC[210]//TiC[110]//Al$_2$O$_3$[210]. The values of the *a*-axis and *c*-axis were determined to be 3.04 and 13.59 Å by reciprocal space mapping (RSM). The epitaxial growth behavior has also been documented by transmission electron microscopy (TEM) [25-29]. XPS-analysis depth profiles of the deposited films within the present study using a PHI Quantum instrument, showed after 60 seconds of Ar-sputtering a constant composition without any contamination species.

**Computational details**

**Calculation of the x-ray emission spectra**
The x-ray emission spectra were calculated within the single-particle transition model by using the APW+lo band structure method [30]. Exchange and correlation effects were described by means of the generalized gradient approximation (GGA) as parameterized by Perdew, Burke and Ernzerhof [31]. A plane wave cut-off, corresponding to $R_{MT}*K_{max}=8$, was used in the present investigation. For Ti, *s* and *p* local orbitals were added to the APW basis set to improve the convergence of the wave function while, for C only *s* local orbitals were added to the basis set. In order to calculate the Al $L_{2,3}$-edge the *1s*, *2s* and *2p* orbitals were treated in Al as core states, leaving therefore only the *3s* and *3p* electrons inside the valence shell. No additional local orbitals were added in this case. The charge density and potentials were expanded up to $l=12$ inside the atomic spheres, and the total energy was converged with respect to the Brillouin zone integration.

The x-ray emission spectra were then evaluated at the converged ground-state density by multiplying the angular momentum projected density of states by a transition-matrix element [32]. The electric-dipole approximation was employed so that only the transitions between the core states with orbital angular momentum $l$ to the $l\pm1$ components of the electronic bands were considered. The core-hole lifetimes used in the calculations were 0.73 eV, 0.27 eV and 0.5 eV for the Ti *2p*, C *1s* and Al *2p* edges, respectively. A direct comparison of the calculated spectra with the measured data was finally achieved by including the instrumental broadening in form of Gaussian functions corresponding to the experimental resolutions (see experimental section IIA). The final state lifetime broadening was accounted for by a convolution with an energy-dependent Lorentzian function with a broadening increasing linearly with the distance from the Fermi level according to the function $a+b(E-E_F)$, where the constants $a$ and $b$ were set to 0.01 eV and 0.05 (dimensionless) [33].

**Balanced crystal orbital overlap population (BCOOP)**
In order to study the chemical bonding of the $Ti_2AlC$ compound, we calculated the BCOOP function by using the full potential linear muffin-tin orbital (FPLMTO) method [34]. In these calculations, the muffin-tin radii were kept as large as possible without overlapping one another (Ti=2.3 a.u., Al=2.2 a.u and C=1.6 a.u.). To ensure a well-converged basis set, a double basis with a total of four different $\kappa^2$ values were used. For Ti, we included the *4s*, *4p* and *3d* as valence states. To reduce the core leakage at the sphere boundary, we also treated the *3s* and *3p* core states as semi-core states. For Al, *3s*, *3p* and *3d* were taken as valence states. The resulting basis formed a single, fully hybridizing basis set. This approach has previously proven to give a well-converged basis [35]. For the sampling of the irreducible wedge of the Brillouin zone, we used a special-k-point method [36] and the number of k points were 512 for $Ti_2AlC$ and 216 for $Ti_3AlC_2$ in the self-consistent total energy calculation. In order to speed up the convergence, a Gaussian broadening of 20 mRy widths was associated with each calculated eigenvalue.

## Results

### Ti $L_{2,3}$ x-ray emission

Figure 3 shows Ti $L_{2,3}$ SXE spectra of Ti$_2$AlC excited at 458, 459.9, 463.6 (resonant) and 477 eV (nonresonant) photon energies, corresponding to the $2p_{3/2}$ and $2p_{1/2}$ absorption maxima and nonresonant excitation, respectively. XAS measurements (top, right curves) were used to locate the energies of the absorption peak maxima. For comparison of the spectral shapes, the measured spectra were normalized to unity and plotted on a photon energy scale (top) and a common energy scale (bottom) with respect to the Fermi level ($E_F$) using the measured $2p_{1/2}$ core-level XPS binding energy of 460.3 eV of the Ti$_2$AlC sample.

The Ti $L_{2,3}$ SXE spectra are rather delocalized (wide bands) which makes electronic structure calculations suitable for interpretation of nonresonant spectra. For comparison, calculated Ti $L_{2,3}$ spectra of Ti$_2$AlC, TiC and pure Ti are shown at the bottom of Fig. 3. The calculated spectra consist of the density of states obtained from *ab initio* density-functional theory including dipole matrix elements where the life-time broadening was set to 0.73 eV both for the $2p_{3/2}$ and $2p_{1/2}$ thresholds. To account for the Coster-Kronig process, the calculated spectra were also fitted to the experimental $L_3/L_2$ ratio of 6:1. Furthermore, the spin-orbit splitting was set to the experimental value of 6.2 eV while the *ab initio* value was 5.7 eV. The fitted spectra of Ti$_2$AlC and TiC are generally in good agreement with the experimental results.

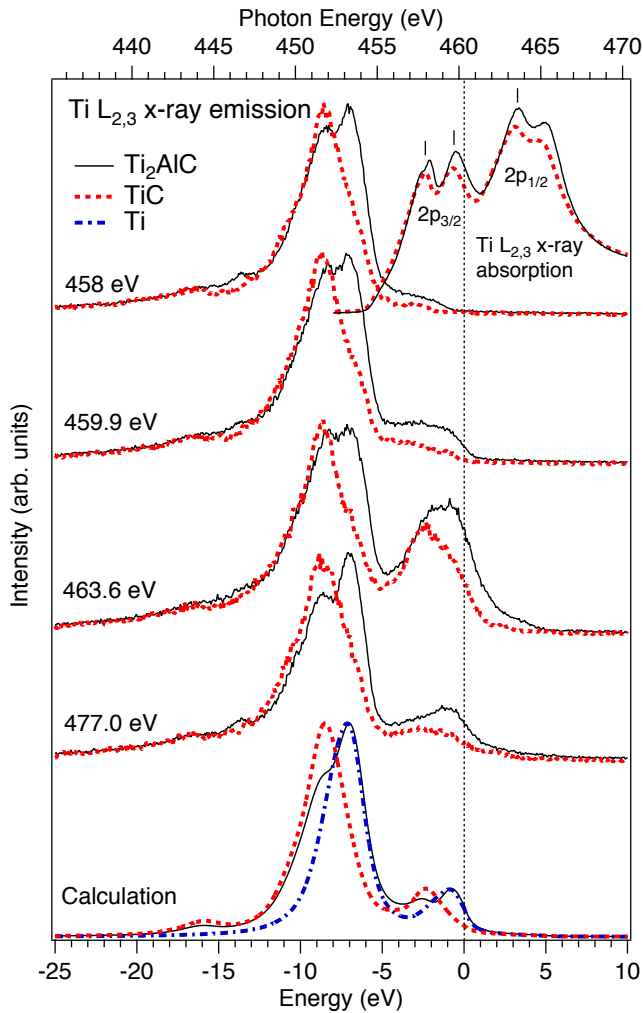

**Fig. 3:** Top, Ti $L_{2,3}$ x-ray emission spectra of Ti$_2$AlC and TiC excited at 458, 459.9, 463.6 and 477 eV. The excitation energies for the resonant emission spectra are indicated by vertical ticks in the x-ray absorption spectra (top, right curves). All spectra are aligned to the Ti $2p_{1/2}$ threshold at 460.3 eV measured by XPS on the Ti$_2$AlC sample. Bottom, fitted spectra with the experimental $L_{2,3}$ peak splitting of 6.2 eV and the $L_3/L_2$ ratio of 6:1 compared to the x-ray emission spectra excited at 477.0 eV.

The main $L_3$ and $L_2$ emission lines are observed at 7 eV and 1 eV on the common energy scale at the bottom. Note that the Ti $L_{2,3}$ SXE spectral shapes of Ti$_2$AlC and TiC are quite different, with part of the main peak coinciding at 8.5 eV, indicating carbide-like attributes. As the excitation energy is changed, the main difference between the spectra is the $L_2$ emission line, which resonates at 463.6 eV, corresponding to the $2p_{1/2}$ absorption maximum. The most significant feature in the Ti SXE spectra of Ti$_2$AlC is the pronounced double peak observed both in the experiment and in the calculation. This double peak has a splitting of 1.5 eV. The origin of the main 7 eV peak is related to a series of flat bands of *3d* character. Note that the 7 eV peak has a significant energy dependence at the $2p_{3/2}$ threshold and does not exist at all in TiC. The double peak is less pronounced at the $2p_{1/2}$ threshold due to the larger core-hole lifetime broadening. Since the 7 eV peak does not exist in systems where Al is replaced by Si and Ge, it is a signature of hybridization between the Ti *3d* states and the Al *3p* states at the top of the valence band. A similar pronounced double peak has been observed in Ti $L_{2,3}$ SXE spectra of Ti$_3$AlC$_2$ with the same peak splitting of 1.5 eV but with much more weight on the 8.5 eV carbide peak [16]. The relative difference between the 7 eV and 8.5 eV peak intensities can be explained by the fact that Ti$_2$AlC, Ti$_3$AlC$_2$ and TiC all contain the same relative amount of Ti atoms (50%) but Ti$_2$AlC also contains 8% more Al and 8% less C than Ti$_3$AlC$_2$ referring to the number of Ti layers over number of all layers in one unit cell. This is a clear indication of two separate contributions with different origins. Comparing the Ti $L_{2,3}$ SXE spectra of Ti$_2$AlC with the parent compound TiC, it is thus possible to understand the changes in the electronic structure when all Al atoms are replaced by C in Ti$_2$AlC. Since the Ti peak at 7 eV completely disappears in TiC, it strongly depends on the relative amount of Al in the system. On the contrary, the carbide peak observed at 8.5 eV is due to the Ti *3d* - C *2p* hybridization. The weak carbide structure observed around 16 eV below $E_F$ is related to Ti *3d* - C *2s* hybridization. In Ti$_2$AlC another weak peak feature is also experimentally identified at 14 eV below $E_F$ but it is not reproduced in the calculation. This feature is either due to an overlap, which is not reproduced theoretically, or due to shake-up satellites in the final state of the x-ray emission process.

Finally, we note that nanolaminated MAX-phases, including Ti$_2$AlC are slightly anisotropic in nature, and therefore exhibit some polarization dependence for the Ti $L_{2,3}$ SXE spectra. We have estimated this effect by estimating in the dipole approximation the effect of matrix elements corresponding to the E-vector of the light parallel to the z-axis and perpendicular to the z-axis. The anisotropy in the calculated spectra is not very pronounced (and for this reason these data are not shown) but we note that the main effect is that there is an enhancement of the 8.5 eV carbide peak, which further improves the agreement with experiment.

**C *K* x-ray emission**
Figure 4 (top) shows experimental C *K* SXE spectra of Ti$_2$AlC and TiC, excited at 284.5 and 285.5 eV (resonant) and 310 eV (nonresonant) photon energies. XAS spectra (top, right curves) were measured to identify the absorption maxima and the excitation energies for the emission spectra. Calculated emission spectra are shown at the bottom of Figure 4. The agreement between the experimental and theoretical spectra is generally good and anisotropic effects are predicted to be small for C *K* SXE. The main peak 2.9 eV below E$_F$ has a shoulder on the low-energy side at 4.0 eV below E$_F$. For resonant excitation, the 4.0-eV shoulder on the low-energy side is more pronounced in Ti$_2$AlC, while for nonresonant excitation it more pronounced in TiC. The TiC spectra indicate what the electronic structure of Ti$_2$AlC would look like if Al would be exchanged to C. Although the C-Al interaction is weak, the spectral differences indicate a more pronounced low-energy shoulder in TiC and more weight towards the E$_F$. In contrast to Ti$_2$AlC, Ti$_3$AlC$_2$ has a high-energy shoulder at 2 eV [16]. The agreement between the experimental and calculated spectra is good although the low-energy shoulder at 4.0 eV is more pronounced in the calculation than what is observed experimentally. The main peak and the shoulder corresponds to the occupied C *2p* orbitals mainly hybridized with the Ti *3d* orbitals of the valence bands with some influence of the Al states.

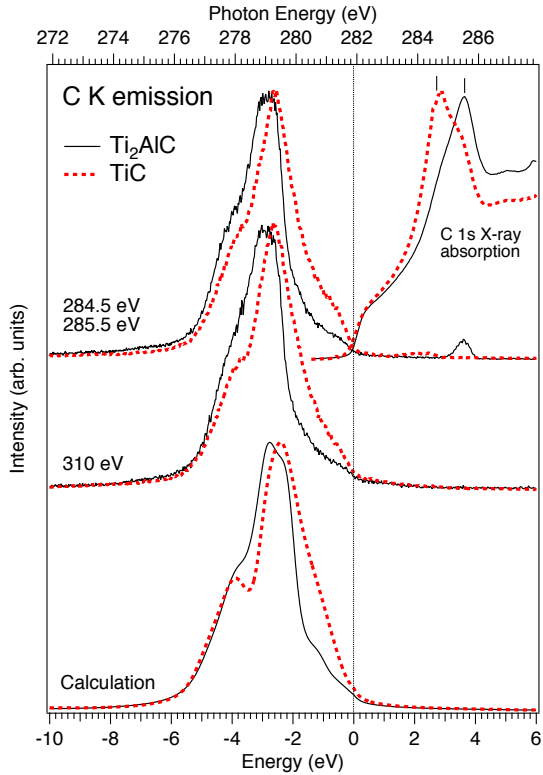

**Fig. 4:** Top, experimental C *K* SXE spectra of Ti$_2$AlC and TiC excited at 284.5, 285.5 (resonant) and 310 (nonresonant) aligned with the C *1s* core XPS binding energy 281.9 eV for Ti$_2$AlC. The resonant excitation energies for the SXE spectra are indicated in the C *1s* XAS spectra (top, right curves) by the vertical ticks. Note the corresponding elastic peak at 285.5 eV in the resonant emission spectrum for Ti$_2$AlC. Bottom, calculated emission spectra of Ti$_2$AlC and TiC. The vertical dotted line indicates the Fermi level (E$_F$).

**Al $L_{2,3}$ x-ray emission**

Figure 5 shows an experimental Al $L_{2,3}$ SXE spectrum of Ti$_2$AlC in comparison to Ti$_3$AlC$_2$ from ref [16] both measured nonresonantly at 120 eV photon energy. Comparing the experimental and calculated spectra, it is clear that the main peak at 3.6 eV below E$_F$ of the SXE spectrum is dominated by *3s* final states. The partly populated *3d* states form the broad peak structure close to E$_F$ and participate in the Ti-Al bonding in Ti$_2$AlC. As observed, the Al $L_{2,3}$ SXE spectrum of Ti$_2$AlC has fewer sub-structures than Ti$_3$AlC$_2$ [16]. This shows that the Ti *3d* - Al *3p* hybridization is different in Ti$_2$AlC than in Ti$_3$AlC$_2$ in the energy region 2 to 4 eV below E$_F$. Since the Al *3p* states dominate in the upper part of the Al $L_{2,3}$ valence band, their hybridization indirectly contribute to the spectral shape of the Al $L_{2,3}$ SXE spectra although they are dipole forbidden. For the Al $L_{2,3}$ SXE spectrum, the valence-to-core matrix elements are found to play an important role to the spectral shape. In contrast to Al $L_{2,3}$ SXE spectra of pure Al, which have a sharp and dominating peak structure 1 eV below E$_F$, the Al $L_{2,3}$ SXE spectrum of Ti$_2$AlC has a strongly modified spectral weight towards lower energy. A similar modification of the Al $L_{2,3}$ SXE spectral shape has been observed in the metal aluminides [37]. Comparing the spectral shape to the aluminides, the appearance of the broad low-energy structure around 5.5 eV below $E_F$ in the Al $L_{2,3}$ SXE spectrum of $Ti_2AlC$ can be attributed to the formation of hybridized Al *3s* states produced by the overlap of the Ti *3d*-orbitals. This interpretation is supported by our first principle calculations. The anisotropy (polarization dependence) of Al $L_{2,3}$ SXE spectra of $Ti_2AlC$ is expected to be small due to the dominating *3s* contribution with spherical symmetry.

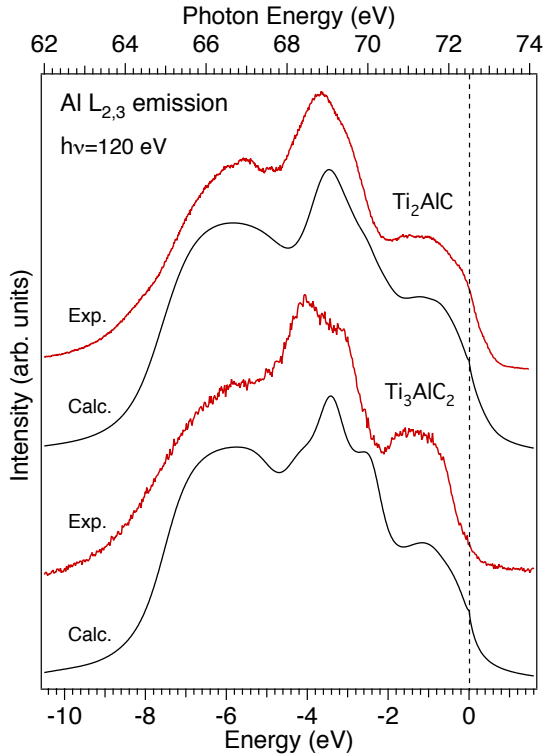

**Fig. 5:** Experimental and calculated Al $L_{2,3}$ SXE spectra of $Ti_2AlC$ compared to those of $Ti_3AlC_2$ from ref [16]. The vertical dotted line indicates the Fermi level ($E_F$).

**Chemical Bonding**
By relaxing the cell parameters of $Ti_2AlC$, the calculated equilibrium *a* and *c*-axis were obtained. For $Ti_2AlC$, they were determined to be 3.08 Å and 13.77 Å, respectively. These values are in good agreement with the experimental values of 3.04 and 13.59 Å presented in section IIB. In order to analyze the chemical bonding in more detail, we show in Fig. 6 the calculated BCOOP [38] of $Ti_2AlC$ compared to $Ti_3AlC_2$ [16] and TiC. The BCOOP makes it possible to compare the strength of two similar chemical bonds and is a positive function for bonding states and negative for anti-bonding states. The strength of the covalent bonding can be determined by comparing the areas under the BCOOP curves. The energy distance position of the peaks from the $E_F$ also gives an indication of the strength of the covalent bonding. Firstly, comparing the areas under the BCOOP curves and the distances of the main peaks of the curves from the $E_F$, it is clear that the Ti *3d* - C *2p* bond is much stronger than the Ti *3d* - Al *3p* bond in both $Ti_2AlC$ and $Ti_3AlC_2$. The Ti atoms bond more strongly to C than Al, which gives rise to a stronger Ti-C bond for $Ti_{II}$ than for $Ti_I$ in the case of $Ti_3AlC_2$. Consequently, the Ti-C chemical bond is stronger in $Ti_2AlC$ than in TiC as shown by the shorter bond length in Table I. Secondly, comparing the BCOOP curves of $Ti_2AlC$ to those of $Ti_3AlC_2$ and TiC, the Ti-C BCOOP curve of $Ti_2AlC$ is the most intense which indicates that the Ti-C bond is slightly stronger in $Ti_2AlC$ than in $Ti_3AlC_2$ and TiC. For the Ti $L_{2,3}$ SXE spectrum of $Ti_2AlC$, discussed in section IVA, the BCOOP calculations confirm that the Ti *3d* - C *2p* hybridization and strong covalent bonding is in fact the origin of the low-energy carbide peak at 8.8 eV below the $E_F$ (2.6 eV in Fig. 6 when the spin-obit splitting is not taken into account). Although a single carbide peak is observed experimentally, the BCOOP analysis show that there are several overlapping energy levels in the region between 2.0 and 5.5 eV below $E_F$. Thirdly, the Ti-Al BCOOP peak of $Ti_2AlC$ is slightly weaker and closer to the $E_F$ than in $Ti_3AlC_2$. This is an indication that the Ti-Al chemical bond in $Ti_2AlC$ is somewhat weaker than in $Ti_3AlC_2$. This is also verified experimentally by the fact that the spectral weight of the Ti $L_{2,3}$ SXE spectrum is slightly shifted towards the $E_F$ which plays a key role for the physical properties.

**TABLE I:** Calculated bond lengths for TiC, $Ti_2AlC$ and $Ti_3AlC_2$. In $Ti_3AlC_2$, $Ti_I$ is connected to C while $Ti_{II}$ is connected to both C and Al as illustrated in Fig. 1.

| Bond type | $Ti_I$ - C | $Ti_{II}$ - C | Al - $Ti_{II}$ | Al - $Ti_I$ | Al - C |
|---|---|---|---|---|---|
| TiC | 2.164 | | | | |
| $Ti_2AlC$ | | 2.117 | 2.901 | | 3.875 |
| $Ti_3AlC_2$ | 2.201 | 2.086 | 2.885 | 4.655 | 3.802 |

Our Ti $2p_{3/2,1/2}$ core-level XPS values of the $Ti_2AlC$ sample (454.2 eV and 460.3 eV, respectively), show that there is a high-energy shift of the binding energies due to screening in comparison to pure Ti (453.8 eV and 460.0 eV, respectively). This is an indication of charge-transfer from Ti to C and Al. On the contrary, the XPS-binding energies of Al in $Ti_2AlC$ are shifted to lower energy (72.5 eV) in comparison to pure Al (72.8 eV). This is more pronounced for C (281.9 eV) in comparison to amorphous C-C carbon (284.8 eV) although only carbide-type of carbon is relevant here. A similar trend of the chemical shift has been found for the XPS-binding energies in $Ti_3AlC_2$ [39].

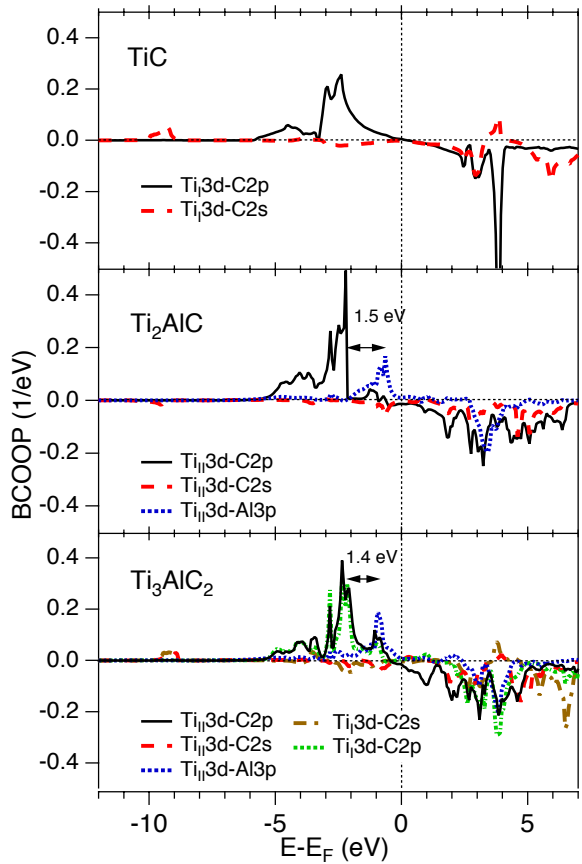

**Fig. 6:** Calculated balanced crystal overlap population (BCOOP) of TiC, $Ti_2AlC$ and $Ti_3AlC_2$. Note that the Ti $3d$ - C $2s$ overlap around 10 eV below $E_F$ is antibonding in $Ti_2AlC$ and bonding for $Ti_3AlC_2$ and TiC. The $Ti_I$ and $Ti_{II}$ atoms have different chemcial environments as shown in Fig. 1.

Figure 7 shows a calculated electron density difference plot between $Ti_2AlC$ and $Ti_2C_2$, where in the latter Al has been replaced by C in the same 211 crystal structure representing a highly twisted TiC structure i.e., $Ti_2C_2$. The plot was obtained by taking the difference between the charge densities of the two systems in the [110] planes of the hexagonal unit cell. When introducing the Al atoms into the $Ti_2C_2$ crystal structure we first observe an anisotropic charge variation around the Ti atoms. In particular, in the direction along the Ti-Al bond (~45° angle to the corners of the plot) we register an electron density withdrawal (see the red area around Ti) from Ti to Al as to indicate the formation of the Ti-Al bonds. The consequence of such an electronic movement is the creation of a certain polarization on the neighbor Ti-Ti bonding and therefore to reduce its strength. The insertion of the Al atoms in the $Ti_2C_2$ structure locally introduce an anisotropic electron density distribution around the Ti atoms resulting in a whole charge-modulation along the Ti-Al-Ti-Ti-Al-Ti zigzag bonding direction that propagates throughout the unit cell. The charge transfer from Ti towards Al is in agreement with our measured XPS core-level shifts and the BCOOP presented in Fig. 6. Finally, we also observe that the charge-density difference is zero at the carbon atoms at the corners of the plot in Fig. 7. This is an indication that the carbon atoms do not respond markedly to the introduction of Al planes and implies that Al substitution only results in local modifications to the charge density, and possibly a weak Al-C interaction. A very weak Al-C bond has also been presented experimentally [40].

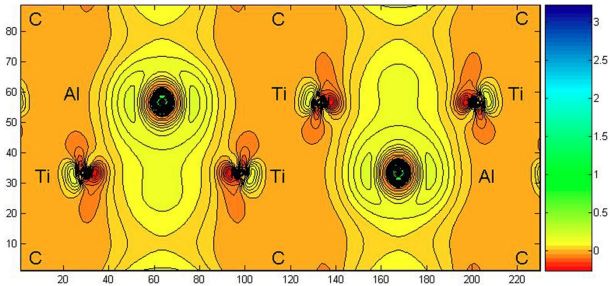

**Fig. 7:** Calculated electron density difference plot between $Ti_2AlC$ and $Ti_2C_2$ (TiC) in the same crystal geometry. A carbon atom is located in each corner of the plot where the charge-density difference is zero. The difference density plot was obtained by subtracting the charge densities in the [110] diagonal plane of the hexagonal unit cell. The lower valence band energy was fixed to -1.0 Ry (-13.6 eV) and all the Ti *3d*, *4s*; Al *3s*, *3p* and C *2s 2p* valence states were taken into account.

**Discussion**

Comparing the crystal structure of $Ti_2AlC$ in Fig. 1 with those of $Ti_3AlC_2$ and TiC, it is clear that the physical properties and the underlying electronic structure of the Ti-Al-C system is strongly affected by the number of Al layers per Ti layer. In $Ti_2AlC$, there is one Al layer for every second layer of Ti while in $Ti_3AlC_2$ there is one Al layer for each third Ti layer. In $Ti_3AlC_2$ there are two types of Ti sites ($Ti_I$ and $Ti_{II}$) while only one Ti site exist in $Ti_2AlC$ and TiC. The Ti SXE spectra in Fig. 3 show that the intensity at the $E_F$ is considerably higher in $Ti_2AlC$ than in TiC. This is also the case for $Ti_3AlC_2$ [16]. For C in Fig. 4, the intensity at the $E_F$ is similar for both $Ti_2AlC$ and TiC. For Al in Fig. 5, the intensity at the $E_F$ is higher in $Ti_2AlC$ than in $Ti_3AlC_2$. Intuitively, one would therefore expect that the conductivity would increase as more Al monolayers are introduced since Al metal is a good conductor. However, in $Ti_2AlC$, the $E_F$ is close to a pronounced pseudogap (a region with low density of states) of the dominating Ti *3d* states. The conductivity is largely governed by the Ti metal bonding and is roughly proportional to the number of states at the Fermi level (TiC: 0.12 states/eV/atom, $Ti_2AlC$: 0.34 states/eV/atom and $Ti_3AlC_2$: 0.33 states/eV/atom). The $Ti_2AlC$ ternary carbide film thus has a similar resistivity (0.4 µΩ m) compared to $Ti_3AlC_2$ (0.5 µΩ m. In our previous 312 study [16], it was clear that the $Ti_{II}$ layers contribute more to the conductivity than the $Ti_I$ layers. Therefore, one would also expect that $Ti_2AlC$ has higher conductivity than all 312-phases since it only contains $Ti_{II}$. The states near $E_F$ are dominated by Ti *3d* orbitals with contribution from Al *3p* orbitals. However, the metal-metal *dd* interactions (metal bonding) play an important role close to $E_F$ and the Ti-Al-C MAX-phases show excellent conductivity due to the metallic bonding.

From Figure 3, we identified two types of bonds, the strong Ti *3d* - C *2p* carbide bond and the weaker Ti *3d* - Al *3p* aluminum bond. The Ti *3d* - C *2p* and Ti *3d* - C *2s* hybridizations are both deeper in energy from the $E_F$ than the Ti *3d* - Al *3p* hybridization which is an indication of a stronger bonding. A strengthening of the relatively weak covalent Ti *3d* - Al *3p* bonding effectively increase the shear stiffness (hardness and elasticity). This is observed in $Ti_2AlC$ in comparison to $Ti_3AlC_2$ as the E-modulus increases with decreasing number of Al layers per Ti layer, from 240 GPa to 260 GPa. The E-modulus of both $Ti_2AlC$ and $Ti_3AlC_2$ is lower than for TiC (350-400 GPa). The softening of the $Ti_2AlC$ is due to changes in the bonding conditions of the weaker Ti-Al bonds. In this sense, $Ti_3AlC_2$ show more carbide-like attributes and is more similar to TiC than $Ti_2AlC$ since there is a reduced number of inserted Al monolayers. The deformation and delamination mechanism is similar in both systems due to the weak Ti-Al bonds. Our results show clear differences between the electronic structures of the two MAX phases. The properties of the Ti-Al-C systems are thus directly related to the number of inserted Al layers into the TiC matrix. This is due to the weak covalent bond between Ti and Al compared to Ti-C, which softens the material. By tuning the Al content, the physical and mechanical properties can thus be custom made for specific applications.

**Conclusions**
In summary, we have investigated the electronic structure of $Ti_2AlC$ and compared the results to those of TiC and $Ti_3AlC_2$ with the combination of soft x-ray emission spectroscopy and electronic structure calculations. The origin of a pronounced double-peak structure in Ti $L_{2,3}$ x-ray emission is identified having different spectral intensity weights in $Ti_2AlC$ and $Ti_3AlC_2$. The peak structure observed 2.6 eV below the Fermi level is shown to be due to Ti *3d* - C *2p* hybridization and strong covalent bonding while another peak observed 1 eV below the Fermi level is due to carbide Ti *3d* states with hybridization with Al *3p* states with a weaker covalent bonding. In addition, carbide Ti *3d* - C *2s* hybridization is identified around 10 eV below the Fermi level as a weak spectral structure in Ti $L_{2,3}$ emission. The calculated orbital overlaps indicate that the Ti *3d* - Al *3p* bonding orbitals of $Ti_2AlC$ are somewhat weaker than in $Ti_3AlC_2$ which implies a change of the elastic properties and the electrical and thermal conductivity. The analysis of the underlying electronic structure thus provides increased understanding of the difference of materials properties between $Ti_2AlC$, $Ti_3AlC_2$ and TiC. As in the case of $Ti_3AlC_2$, the Al $L_{2,3}$ x-ray emission spectra of Al in $Ti_2AlC$ appear very different from the pure Al metal indicating strong hybridization between the A-atoms with Ti. Generally, the covalent bonding mechanism is very important for the mechanical and physical properties of these thermodynamically stable nanolaminates. A tuning of the physical and mechanical properties by insertion of more or fewer Al layers in the TiC matrix implies that these nanolaminated carbide systems can be custom-made by the choice of phase or composition by changing the number of interleaved Al layers in TiC

**Acknowledgements**
We would like to thank the staff at MAX-lab for experimental support. This work was supported by the Swedish Research Council, the Göran Gustafsson Foundation, the Swedish Strategic Research Foundation (SSF) Materials Research Programs on Low-Temperature Thin Film Synthesis and the Swedish Agency for Innovation Systems (VINNOVA) Project on Industrialization of MAX Phase Coatings.